\documentclass[twocolumn,showpacs,preprintnumbers,amsmath,amssymb]{revtex4}

\usepackage{graphicx}
\usepackage{dcolumn}
\usepackage{bm}

\newcommand{\bq}{\begin{equation}}
\newcommand{\eq}{\end{equation}}
\newcommand{\bqa}{\begin{eqnarray}}
\newcommand{\eqa}{\end{eqnarray}}
\newcommand{\nn}{\nonumber \\}
\newcommand{\ij}{\langle i j \rangle}

\def\be     {\begin{equation}}
\def\ee     {\end{equation}}
\def\bea        {\begin{eqnarray}}
\def\eea        {\end{eqnarray}}
\def\bnn    {\begin{eqnarray*}}
\def\enn    {\end{eqnarray*}}

\begin{document}

\title{SU(2) gauge theory of the Hubbard model: Emergence of an anomalous metallic phase near the Mott critical
point}
\author{Ki-Seok Kim}
\affiliation{ School of Physics, Korea Institute for Advanced
Study, Seoul 130-012, Korea }
\date{\today}

\begin{abstract}
We propose one possible mechanism for an anomalous metallic phase
appearing frequently in two spatial dimensions, that is, local
pairing fluctuations. Introducing a pair-rotor representation to
decompose bare electrons into collective pairing excitations and
renormalized electrons, we derive an SU(2) gauge theory of the
Hubbard model as an extended version of its U(1) gauge
theory\cite{Florens,LeeLee} to allow only local density
fluctuations. Since our effective SU(2) gauge theory admits two
kinds of collective bosons corresponding to pair excitations and
density fluctuations respectively, an intermediate phase appears
naturally between the spin liquid Mott insulator and Fermi liquid
metal of the U(1) gauge theory,\cite{Florens,LeeLee} characterized
by softening of density-fluctuation modes as the Fermi liquid, but
gapping of pair-excitation modes. We show that this intermediate
phase is identified with an anomalous metallic phase because there
are no electron-like quasiparticles although it is metallic.
\end{abstract}


\maketitle

It has been a common belief for a long time in the condensed
matter community that a metallic phase does not exist at zero
temperature in two spatial dimensions owing to localization by
disorder, even if electron-electron interactions
exist.\cite{Review1,Review2} However, if one considers a non-Fermi
liquid phase instead of a Fermi liquid state, characterized by the
absence of electron quasiparticles due to an anomalous exponent in
the electron spectral function, he can show that disorder-induced
interactions in the replica formalism are irrelevant for moderate
values of the exponent in the renormalization group
sense.\cite{NFL_exponent,Kim_exponent} In this paper we
demonstrate how the non-Fermi liquid metal can arise from strong
electron-electron interactions, based on the Hubbard model as a
minimal model. We show that local pairing fluctuations can cause
an anomalous metal, described by an effective U(1) gauge theory.

One may introduce a local pairing order parameter by utilizing a
Hubbard-Stratonovich (HS) transformation. Integrating out gapless
electrons, one obtains an effective action of the pairing order
parameter with dissipation that results from the gapless electrons
near the Fermi surface. It was conjectured that the presence of
dissipation may result in a metallic phase in two dimensional
interacting electrons.\cite{Review2,Dissipation_Metal} However,
there are several unsatisfactory points in the order parameter
approach. Since the gradient expansion used in this method is
basically a weak coupling approach, the breakdown of this weak
coupling approach is expected in a large coupling limit. In
addition, it is difficult to justify the gradient expansion in the
presence of gapless electrons because they can cause nonlocal
interactions between order parameters, making it unreliable a
conventional treatment in a local effective
action.\cite{Gradient_expansion} Furthermore, the conjecture is
not convincing yet that dissipation can give rise to a metallic
phase in two dimensions.\cite{Review2}

In this respect we utilize a strong coupling approach, that is,
decomposing bare electrons $c_{\sigma}$ into collective
excitations $U_{\sigma\sigma'}$ and renormalized electrons
$f_{\sigma}$, i.e., $c_{\sigma} =
U^{\dagger}_{\sigma\sigma'}f_{\sigma'}$.\cite{Slave_boson} Here
local pairing excitations can be imposed in the
$U_{\sigma\sigma'}$ field. Recently, Florens and George proposed a
slave-rotor representation of the Hubbard model, where the
collective boson mode is set to be $U_{\sigma\sigma'} =
e^{i\theta}\delta_{\sigma\sigma'}$, associated with local density
fluctuations.\cite{Florens} Using the slave-rotor representation,
Lee and Lee constructed a U(1) gauge theory of the Hubbard model
to study a spin liquid Mott insulator to Fermi liquid transition
in triangular lattice.\cite{LeeLee} In this paper we extends the
U(1) gauge theory formulation by allowing local pairing
fluctuations. Then, the collective boson field $U_{\sigma\sigma'}$
is expressed as an SU(2) matrix field, involved with both density
and pairing fluctuations. This leads us to construct an SU(2)
gauge theory of the Hubbard model. Since our SU(2) gauge theory
admits two kinds of collective bosons corresponding to density
fluctuations and local pairing excitations respectively, an
intermediate phase is generically allowed between the spin liquid
Mott insulator and Fermi liquid metal of the slave-rotor U(1)
gauge theory via SU(2) symmetry breaking due to hole doping. We
show that this intermediate state is identified with an anomalous
metallic phase in the respect that there are no electron-like
quasiparticles although it is metallic.

We consider the Hubbard Hamiltonian \bqa && H = -
t\sum_{ij\sigma}c_{i\sigma}^{\dagger}c_{j\sigma} +
\frac{3u}{2}\sum_{i}c_{i\uparrow}^{\dagger}c_{i\uparrow}c_{i\downarrow}^{\dagger}c_{i\downarrow}
, \eqa where $t$ is a hopping integral of electrons, and $u$
strength of on-site Coulomb repulsions. The local interaction term
can be decomposed into pairing and density channels in the
following way \bqa &&
\frac{3u}{2}\sum_{i}c_{i\uparrow}^{\dagger}c_{i\uparrow}c_{i\downarrow}^{\dagger}c_{i\downarrow}
=
\frac{u}{2}\sum_{i}c_{i\uparrow}^{\dagger}c_{i\downarrow}^{\dagger}
c_{i\downarrow}c_{i\uparrow} \nn && +
\frac{u}{2}\sum_{i}\Bigl(\sum_{\sigma}c_{i\sigma}^{\dagger}c_{i\sigma}-1\Bigr)^{2}
+
\frac{u}{2}\Bigl(\sum_{\sigma}c_{i\sigma}^{\dagger}c_{i\sigma}-1\Bigr)
. \nonumber \eqa Performing the HS transformation for the pairing
and density interaction channels, we find an effective Lagrangian
in the Nambu-spinor representation \bqa &&  Z =
\int{D[\psi_{i},\psi_{i}^{\dagger}, \Phi^{R}_{i}, \Phi_{i}^{I},
\varphi_{i}]}e^{-\int{d\tau} L} , \nn && L =
\sum_{i}\psi_{i}^{\dagger}(\partial_{\tau}\mathbf{I} -
\mu\tau_{3})\psi_{i} -
t\sum_{\ij}(\psi_{i}^{\dagger}\tau_{3}\psi_{j} + H.c.) \nn && -
i\sum_{i}(\Phi^{R}_{i}\psi_{i}^{\dagger}\tau_{1}\psi_{i} +
\Phi^{I}_{i}\psi_{i}^{\dagger}\tau_{2}\psi_{i} +
\varphi_{i}\psi_{i}^{\dagger}\tau_{3}\psi_{i}) \nn && +
\frac{1}{2u}\sum_{i}(\Phi^{R2}_{i} + \Phi^{I2}_{i} +
\varphi_{i}^{2}) . \eqa Here $\psi_{i}$ is the Nambu spinor, given
by $\psi_{i}= \left(\begin{array}{c} c_{i\uparrow} \\
c_{i\downarrow}^{\dagger} \end{array} \right)$.  $\Phi^{R}_{i}$
and $\Phi^{I}_{i}$ are the real and imaginary parts of the
superconducting order parameter respectively, and $\varphi_{i}$ an
effective density potential. $\mu$ is an electron chemical
potential which differs from its bare value $\mu_{b}$ as $\mu =
\mu_{b} - u/2$. Introducing a pseudospin vector $\vec{\Omega}_{i}
\equiv (\Phi_{i}^{R}, \Phi_{i}^{I}, \varphi_{i})$, one can express
Eq. (2) in a compact form \bqa && Z =
\int{D[\psi_{i},\psi_{i}^{\dagger},
\vec{\Omega}_{i}]}e^{-\int{d\tau} L} , \nn && L =
\sum_{i}\psi_{i}^{\dagger}(\partial_{\tau}\mathbf{I} -
\mu\tau_{3})\psi_{i} -
t\sum_{\ij}(\psi_{i}^{\dagger}\tau_{3}\psi_{j} + H.c.) \nn && -
i\sum_{i}\psi_{i}^{\dagger}(\vec{\Omega}_{i}\cdot\vec{\tau})\psi_{i}
+
\frac{1}{4u}\sum_{i}\mathbf{tr}(\vec{\Omega}_{i}\cdot\vec{\tau})^{2}
. \eqa Integrating over the pseudospin field $\vec{\Omega}_{i}$,
Eq. (3) recovers the Hubbard model Eq. (1).

As discussed in the introduction, we disintegrate bare electrons
into collective excitations and renormalized electrons in the
following way \bqa \psi_{i} & = &
e^{-i\phi_{1i}\tau_{1}-i\phi_{2i}\tau_{2}-i\phi_{3i}\tau_{3}}\eta_{i}
\equiv U_{i}^{\dagger}\eta_{i} , \eqa  where the two component
spinor $\eta_{i} = \left( \begin{array}{c} \eta_{i+} \\
\eta_{i-}^{\dagger} \end{array} \right)$ can be considered to
express renormalized electrons, and the SU(2) matrix field $U_{i}
= \exp[i\sum_{k=1}^{3}\phi_{ki}\tau_{k}]$ collective bosons. Here
$\exp[i\phi_{1i}\tau_{1}]$ (or $\exp[i\phi_{2i}\tau_{2}]$) can be
interpreted as a creation operator of an electron pair since it
mixes a particle with a hole while $\exp[i\phi_{3i}\tau_{3}]$ is
identified with a creation operator of an electron charge,
corresponding to the rotor variable in the slave-rotor
representation.\cite{Florens,Kim1}

Inserting the decomposition Eq. (4) into the effective Lagrangian
Eq. (3), we obtain \bqa && Z = \int{D[\eta_{i},\eta_{i}^{\dagger},
U_{i}, \vec{\Omega}_{i}]} e^{-\int{d\tau} L} , \nn && L =
\sum_{i}\eta_{i}^{\dagger}(\partial_{\tau}\mathbf{I} +
U_{i}\partial_{\tau}U_{i}^{\dagger}-
\mu{U}_{i}\tau_{3}U_{i}^{\dagger})\eta_{i} \nn && -
t\sum_{\ij}(\eta_{i}^{\dagger}U_{i}\tau_{3}U_{j}^{\dagger}\eta_{j}
+ H.c.) \nn && -
i\sum_{i}\eta_{i}^{\dagger}U_{i}(\vec{\Omega}_{i}\cdot\vec{\tau})U_{i}^{\dagger}\eta_{i}
+
\frac{1}{4u}\sum_{i}\mathbf{tr}(\vec{\Omega}_{i}\cdot\vec{\tau})^{2}
. \eqa Performing the gauge transformation
$\vec{\Omega}_{i}\cdot\vec{\tau} \rightarrow
U_{i}^{\dagger}(\vec{\Omega}_{i}\cdot\vec{\tau})U_{i}$, and
shifting $\vec{\Omega}_{i}\cdot\vec{\tau} \rightarrow
\vec{\Omega}_{i}\cdot\vec{\tau} - i
U_{i}\partial_{\tau}U_{i}^{\dagger} +
i\mu{U}_{i}\tau_{3}U_{i}^{\dagger}$, Eq. (5) reads \bqa && Z =
\int{D[\eta_{i},\eta_{i}^{\dagger}, U_{i}, \vec{\Omega}_{i}]}
e^{-\int{d\tau} L} , \nn && L =
\sum_{i}\eta_{i}^{\dagger}(\partial_{\tau}\mathbf{I} -
i\vec{\Omega}_{i}\cdot\vec{\tau})\eta_{i} -
t\sum_{\ij}(\eta_{i}^{\dagger}U_{i}\tau_{3}U_{j}^{\dagger}\eta_{j}
+ H.c.) \nn && +
\frac{1}{4u}\sum_{i}\mathbf{tr}(-iU_{i}\partial_{\tau}U_{i}^{\dagger}
+ \vec{\Omega}_{i}\cdot\vec{\tau} +
i\mu{U}_{i}\tau_{3}U_{i}^{\dagger})^{2} . \eqa

Using the HS transformation for the hopping term \bqa && -
t(\eta_{i\alpha}^{\dagger}U_{i\alpha\beta}\tau_{3\beta\gamma}U_{j\gamma\delta}^{\dagger}\eta_{j\delta}
+ H.c.) \nn && \rightarrow
t\Bigl[F_{ij\alpha\delta}E_{ij\delta\alpha}^{\dagger} +
E_{ij\alpha\delta}F_{ij\delta\alpha}^{\dagger} \nn && -
(\eta_{i\alpha}^{\dagger}F_{ij\alpha\delta}\eta_{j\delta} +
U_{i\alpha\beta}\tau_{3\beta\gamma}U_{j\gamma\delta}^{\dagger}E_{ij\delta\alpha}^{\dagger})
- H.c. \Bigr] , \nonumber \eqa we find an effective Lagrangian of
the Hubbard model \bqa && Z = \int{D[\eta_{i},\eta_{i}^{\dagger},
U_{i}, \vec{\Omega}_{i}, E_{ij}, F_{ij}]}e^{-\int{d\tau} L} , \nn
&& L = L_{0} + L_{\eta} + L_{U} , \nn && L_{0} =
t\sum_{\ij}\mathbf{tr}(F_{ij}E_{ij}^{\dagger} +
E_{ij}F_{ij}^{\dagger}) , \nn && L_{\eta} =
\sum_{i}\eta_{i}^{\dagger}(\partial_{\tau}\mathbf{I} -
i\vec{\Omega}_{i}\cdot\vec{\tau})\eta_{i} -
t\sum_{\ij}(\eta_{i}^{\dagger}F_{ij}\eta_{j} + H.c.) , \nn &&
L_{U} =
\frac{1}{4u}\sum_{i}\mathbf{tr}(-iU_{i}\partial_{\tau}U_{i}^{\dagger}
+ \vec{\Omega}_{i}\cdot\vec{\tau} +
i\mu{U}_{i}\tau_{3}U_{i}^{\dagger})^{2}  \nn && -
t\sum_{\ij}\mathbf{tr}(U_{j}^{\dagger}E_{ij}^{\dagger}U_{i}\tau_{3}
+ H.c.) , \eqa where $E_{ij}$ and $F_{ij}$ are HS matrix fields
associated with hopping of $\eta_{i}$ fermions and $\phi_{ki}$
bosons, respectively.

We make an ansatz for the hopping matrix fields \bqa && E_{ij}
\approx E\exp[i\vec{a}_{ij}\cdot\vec{\tau}]\tau_{3} , ~~~ F_{ij}
\approx F\exp[i\vec{a}_{ij}\cdot\vec{\tau}]\tau_{3} , \eqa where
$E$ and $F$ are longitudinal modes (amplitudes) of the hopping
parameters, and $\vec{a}_{ij}$ their transverse modes (phase
fluctuations), considered to be spatial components of SU(2) gauge
fields. The reason why we introduce the $\tau_{3}$ matrix is that
the fermion sector $L_{\eta}$ should recover the original electron
Lagrangian Eq. (3) as the slave-rotor
representation\cite{Florens,Kim1} does. This will be addressed
more below Eq. (12).

Inserting Eq. (8) into Eq. (7), we obtain an effective SU(2) gauge
theory of the Hubbard model for the Mott-Hubbard transition \bqa
&& Z = \int{D[\eta_{i},\eta_{i}^{\dagger}, U_{i},
\vec{\Omega}_{i}, \vec{a}_{ij}]}e^{-\int{d\tau} L} , \nn && L =
L_{\eta} + L_{U} + 4t\sum_{\ij} EF , \nn && L_{\eta} =
\sum_{i}\eta_{i}^{\dagger}(\partial_{\tau}\mathbf{I} -
i\vec{\Omega}_{i}\cdot\vec{\tau})\eta_{i} \nn && -
tF\sum_{\ij}(\eta_{i}^{\dagger}e^{i\vec{a}_{ij}\cdot\vec{\tau}}\tau_{3}\eta_{j}
+ H.c.) , \nn && L_{U} =
\frac{1}{4u}\sum_{i}\mathbf{tr}(-iU_{i}\partial_{\tau}U_{i}^{\dagger}
+ \vec{\Omega}_{i}\cdot\vec{\tau} +
i\mu{U}_{i}\tau_{3}U_{i}^{\dagger})^{2}  \nn && -
tE\sum_{\ij}\mathbf{tr}(U_{j}^{\dagger}\tau_{3}e^{-i\vec{a}_{ij}\cdot\vec{\tau}}U_{i}\tau_{3}
+ H.c.) . \eqa In the mean field approximation ignoring SU(2)
gauge fluctuations $\vec{a}_{ij}$, the amplitudes of the hopping
parameters and the SU(2) pseudospin order parameters are
determined by \bqa && 4E =
\langle\eta_{i}^{\dagger}\tau_{3}\eta_{j} + H.c. \rangle , ~~~ 4F
= \langle\mathbf{tr}({U}_{j}^{\dagger}\tau_{3}U_{i}\tau_{3} +
H.c.)\rangle , \nn && {\vec \Omega}_{i} =
iu\langle\eta_{i}^{\dagger}\vec{\tau}\eta_{i}\rangle +
\frac{1}{2}\langle\mathbf{tr}[\vec{\tau}(-iU_{i}\partial_{\tau}U_{i}^{\dagger}
+ i\mu{U}_{i}\tau_{3}U_{i}^{\dagger})]\rangle . \eqa

If one considers an easy-axis limit
$\vec{\Omega}_{i}\cdot\vec{\tau} \equiv \varphi_{i}\tau_{3}$,
$U_{i} \equiv \exp[i\phi_{3i}\tau_{3}]$, and ${\vec
a}_{ij}\cdot{\vec \tau} \equiv a_{3ij}\tau_{3}$, Eq. (9) is
reduced to the effective U(1) gauge Lagrangian in the slave-rotor
representation \bqa && L_{\eta} =
\sum_{i}\eta_{i}^{\dagger}(\partial_{\tau}\mathbf{I} -
i\varphi_{i}\tau_{3})\eta_{i} \nn && -
tF\sum_{\ij}(\eta_{i}^{\dagger}e^{i{a}_{3ij}\tau_{3}}\tau_{3}\eta_{j}
+ H.c.) , \nn && L_{U} =
\frac{1}{2u}\sum_{i}(\partial_{\tau}\phi_{3i} - \varphi_{i} -
i\mu)^{2} \nn && - 2tE\sum_{\ij}\cos(\phi_{3j} - \phi_{3i} -
a_{3ij}) . \eqa One can perform a mean field
analysis\cite{Florens} to show that there exists a
coherent-incoherent transition of $\phi_{3i}$ fields at half
filling ($\varphi_{i} = \mu = 0$), identified with the
Mott-Hubbard transition from a spin liquid Mott insulator to a
Fermi liquid metal.

On the other hand, considering an easy-plane limit
$\vec{\Omega}_{i}\cdot\vec{\tau} \equiv \Phi_{i}^{R}\tau_{1}$,
$U_{i} \equiv \exp[i\phi_{1i}\tau_{1}]$, and ${\vec
a}_{ij}\cdot{\vec \tau} \equiv a_{1ij}\tau_{1}$, we find another
effective U(1) gauge Lagrangian\cite{Canonical_method} \bqa &&
L_{\eta} = \sum_{i}\eta_{i}^{\dagger}(\partial_{\tau}\mathbf{I} -
i\Phi^{R}_{i}\tau_{1})\eta_{i} \nn && -
tF\sum_{\ij}(\eta_{i}^{\dagger}e^{i{a}_{1ij}\tau_{1}}\tau_{3}\eta_{j}
+ H.c.) , \nn && L_{U} =
\frac{1}{2u}\sum_{i}(\partial_{\tau}\phi_{1i} - \Phi_{i}^{R})^{2}
\nn && - 2tE\sum_{\ij}\cos(\phi_{1j}+\phi_{1i} - a_{1ij}) . \eqa
We note that Eq. (12) can be reduced to Eq. (3) with
$\vec{\Omega}_{i}\cdot\vec{\tau} \equiv \Phi_{i}^{R}\tau_{1}$
after the gauge transformations of $\Phi_{i}^{R} \rightarrow
\Phi_{i}^{R} + \partial_{\tau}\phi_{1i}$, $a_{1ij} \rightarrow
a_{1ij} + \phi_{1i} + \phi_{1j}$, and Eq. (4) without $\phi_{2i}$
and $\phi_{3i}$ are utilized. If the $\tau_{3}$ matrix is not
utilized in Eq. (8), the hopping term in $L_{U}$ vanishes, and Eq.
(3) cannot be recovered from Eq. (12). Ignoring gauge
fluctuations, and replacing $\Phi_{i}^{R}$ with $\Phi_{0}$ as the
mean field approximation, we obtain the superconducting order
parameter given by $\Phi_{0} = iu\langle\eta_{i}^{\dagger}
\tau_{1}\eta_{i}\rangle$. It turns out to be zero because double
occupancy costs too much energy. Thus, there is no phase
transition in the fermion Lagrangian as the case of the easy-axis
limit.

To examine the boson Lagrangian in the mean field level, we resort
to large $N$ generalization, following the same method in the
slave-rotor representation.\cite{Florens} Introducing the
$N$-component rotor field $Y_{i}$, we rewrite $L_{U}$ in Eq. (12)
as \bqa && L_{Y} =
\frac{1}{2u}\sum_{i}(\partial_{\tau}Y_{i}^{\dagger})(\partial_{\tau}Y_{i})
- tE\sum_{\ij}(Y_{i}^{\dagger}Y_{j}^{\dagger} + Y_{j}Y_{i}) \nn &&
- i\sum_{i}\lambda_{i}(|Y_{i}|^{2} - 1) , \eqa where $\lambda_{i}$
is a Lagrange multiplier field for the rotor constraint. If we
represent Eq. (13) in terms of $Y_{i} = R_{i} + iI_{i}$ and
$Y_{i}^{\dagger} = R_{i} - iI_{i}$, we find the mean field action
of collective pair excitations \bqa && S_{MF} =
\int_{0}^{\beta}{d\tau}\Bigl[\frac{1}{2u}\sum_{i}[(\partial_{\tau}R_{i})^{2}
+ (\partial_{\tau}I_{i})^{2}] \nn && -2tE\sum_{\ij}(R_{i}R_{j} +
I_{i}I_{j}) - i\lambda_{i}\sum_{i}(R_{i}^{2} + I_{i}^{2} -
1)\Bigr] . \eqa One can find that Eq. (14) is exactly the same as
$\int_{0}^{\beta}{d\tau}L_{U}$ in Eq. (11) at half filling in the
mean field level if the large $N$ generalization for $L_{U}$ in
Eq. (11) is performed by introducing the rotor field $X_{i} =
R_{i} + iI_{i}$. Thus, the mean field analysis in Eq. (11) can be
directly applied to Eq. (12).\cite{MF_analysis} This leads us to
conclude that both $\phi_{3i}$ and $\phi_{1i}$ fields are
simultaneously incoherent in $(u/t)>(u/t)_{0}$, and coherent in
$(u/t)<(u/t)_{0}$ at half filling, where $(u/t)_{0}$ is the
critical value for the Mott transition.\cite{Florens}
Fundamentally, the reason why both fields should be coherent
simultaneously at half filling is the presence of the SU(2)
symmetry at half filling. The slave-rotor action should be
symmetric (invariant) under the transformation $\phi_{1i}
\longleftrightarrow \phi_{3i}$ at half filling in the mean field
approximation.

However, if we consider the case away from half filling, we can
obtain a different result. In the zero doping limit we take into
account the following effective rotor actions from Eq. (9) in the
mean field approximation \bqa && L_{CR} =
\frac{1}{2u}\sum_{i}(\partial_{\tau}\phi_{3i} - i\mu)^{2} -
2tE\sum_{\ij}\cos(\phi_{3j} - \phi_{3i}) , \nn && L_{PR} =
\frac{1}{2u}\sum_{i}(\partial_{\tau}\phi_{1i})^{2} -
2tE\sum_{\ij}\cos(\phi_{1j} + \phi_{1i}) . \eqa It is important to
see that the chemical potential plays the role of Berry phase in
the charge-rotor field $e^{i\phi_{3i}}$, given by $\mu =
u\delta$.\cite{Kim1} One may understand this as SU(2) symmetry
breaking due to hole doping. The presence of the SU(2) symmetry
breaking field is expected to allow an intermediate phase, where
only one phase field is condensed. It is well known that the Berry
phase acts as an effective magnetic field for vortices of
$e^{i\phi_{3i}}$ bosons in the dual representation, disturbing
their condensation.\cite{Magnetic_Field} This implies that
condensation of $e^{i\phi_{3i}}$ bosons can survive to
$(u/t)_{\delta}$ larger than the $(u/t)_{0}$ of half filling. As a
result, the phase characterized by $\langle{e^{i\phi_{3i}}}\rangle
\not= 0$ and $\langle{e^{i\phi_{1i}}}\rangle = 0$ is expected to
appear in the parameter range of $(u/t)_{0}<(u/t)<(u/t)_{\delta}$.

A possible phase diagram in the SU(2) gauge theory of the Hubbard
model is shown in Fig. 1, ignoring antiferromagnetic long range
order. Here SL, FL, and NFL represent spin liquid, Fermi liquid,
and non-Fermi liquid, respectively. It should be noted that the
present mean field analysis is applicable near half filling
because the doping dependence in the hopping amplitudes $E$ and
$F$ is not fully accounted in our analysis. In the small $u/t$
limit both the $\phi_{3i}$ and $\phi_{1i}$ bosons would be
condensed, resulting in a conventional Fermi liquid metal. In the
large $u/t$ limit the two bosons would be gapped, thus a spin
liquid Mott insulator is expected to arise since charge
fluctuations are gapped, but spin excitations remain massless. On
the other hand, in the intermediate range of $u/t$ only the
$\phi_{3i}$ bosons can be condensed while the $\phi_{1i}$ bosons
are not as a result of the Berry phase contribution (SU(2)
symmetry breaking) for the $\phi_{3i}$ field due to hole doping.

\begin{figure}
\includegraphics[width=8cm]{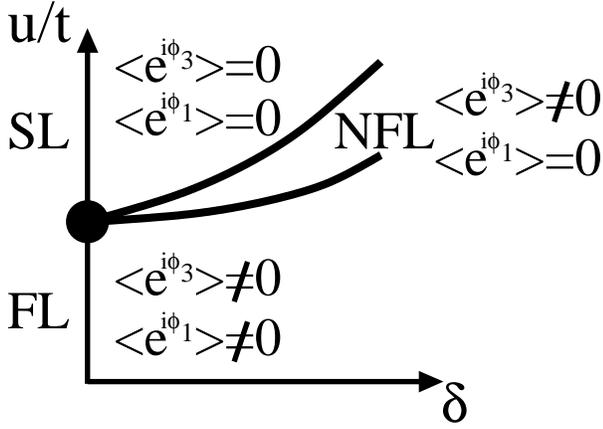}
\caption{\label{Fig. 1} A schematic phase diagram in the SU(2)
gauge theory of the Hubbard model near half filling}
\end{figure}

The intermediate phase is quite interesting in the sense that
although it is a metallic phase, there is no coherent
electron-like quasiparticle peak near the Fermi energy. If one
introduces an external electromagnetic field $A_{ij}$ in Eq. (9),
he can find the following expression of the effective SU(2) gauge
theory in the presence of an electromagnetic field \bqa &&
L_{\eta} = \sum_{i}\eta_{i}^{\dagger}(\partial_{\tau}\mathbf{I} -
i\vec{\Omega}_{i}\cdot\vec{\tau})\eta_{i} \nn && -
tF\sum_{ij}\eta_{i}^{\dagger}e^{i\vec{a}_{ij}\cdot\vec{\tau}}\tau_{3}e^{iA_{ij}\tau_{3}}\eta_{j}
, \nn && L_{U} =
\frac{1}{4u}\sum_{i}\mathbf{tr}(-iU_{i}\partial_{\tau}U_{i}^{\dagger}
+ \vec{\Omega}_{i}\cdot\vec{\tau} +
i\mu{U}_{i}\tau_{3}U_{i}^{\dagger})^{2}  \nn && -
tE\sum_{ij}\mathbf{tr}(U_{j}^{\dagger}e^{-iA_{ij}\tau_{3}}\tau_{3}e^{-i\vec{a}_{ij}\cdot\vec{\tau}}U_{i}\tau_{3}e^{iA_{ij}\tau_{3}})
. \eqa One can check that this gauge action preserves an
electromagnetic U(1) gauge symmetry under the following U(1) gauge
transformations \bqa && \eta_{i} \rightarrow
e^{i\vartheta_{i}\tau_{3}}\eta_{i} , ~~~~~ U_{i} \rightarrow
e^{i\vartheta_{i}\tau_{3}}U_{i}e^{-i\vartheta_{i}\tau_{3}} , \nn
&& e^{i\vec{a}_{ij}\vec{\tau}} \rightarrow
e^{i\vartheta_{i}\tau_{3}}e^{i\vec{a}_{ij}\vec{\tau}}e^{-i\vartheta_{i}\tau_{3}}
, \vec{\Omega}_{i}\cdot\vec{\tau} \rightarrow
e^{i\vartheta_{i}\tau_{3}}\vec{\Omega}_{i}\cdot\vec{\tau}e^{-i\vartheta_{i}\tau_{3}}
, \nn && A_{ij} \rightarrow A_{ij} - (\vartheta_{i} -
\vartheta_{j}) , \eqa where $\vartheta_{i}$ is a U(1) phase angle,
assumed to be independent of time for simplicity. The first line
in Eq. (17) results from the condition $\psi_{i} \rightarrow
e^{i\vartheta_{i}\tau_{3}} \psi_{i}$ in Eq. (4). In the mean field
phase of $\langle{e^{i\phi_{3i}}}\rangle \not= 0$ and
$\langle{e^{i\phi_{1i}}}\rangle = 0$ with $\vec{\Omega}_{i} = 0$
and $\vec{a}_{ij} = 0$, we obtain an effective Lagrangian for the
$\eta_{i}$ fermions from Eq. (16) \bqa && L_{\eta} =
\sum_{i}\eta_{i}^{\dagger}\partial_{\tau}\eta_{i} -
tF\sum_{ij}\eta_{i}^{\dagger}e^{i{A}_{ij}\tau_{3}}\tau_{3}\eta_{j}
. \eqa  This mean field action clearly exhibits a metallic
behavior of the $\eta_{i}$ fermions as the Fermi
liquid.\cite{Gauge_metal} If one ignores the existence of the
$\phi_{1i}$ bosons, this phase is nothing but the Fermi liquid
metal in the slave-rotor theory. However, there exists an
important difference between this metallic phase and Fermi liquid
owing to the gapped $\phi_{1i}$ bosons; there are no electron-like
quasiparticles in the present metallic phase.

One can obtain the following expression for the electron
propagator in our composite field representation \bqa &&
G_{el\uparrow\uparrow}(ij,\tau\tau') =
\langle{T}_{\tau}[c_{\uparrow{i}\tau}c_{\uparrow{j}\tau'}^{\dagger}]\rangle
\nn && =
\langle{T}_{\tau}[e^{-i\phi_{1i\tau}\tau_{1}}\eta_{i\tau}\eta_{j\tau'}^{\dagger}e^{i\phi_{1j\tau'}\tau_{1}}]_{11}\rangle
\nn && =
\langle{T}_{\tau}[\cos\phi_{1i\tau}\cos\phi_{1j\tau'}\eta_{i\tau+}\eta_{j\tau'+}^{\dagger}]\rangle
\nn && +
\langle{T}_{\tau}[\sin\phi_{1i\tau}\sin\phi_{1j\tau'}\eta_{i\tau-}^{\dagger}\eta_{j\tau'-}]\rangle
. \eqa Using the large $N$ generalization in Eq. (14), and
performing the mean field approximation, the electron propagator
is expressed as the convolution of $\eta_{i}$ and $\phi_{1i}$
propagators \bqa && G_{el\uparrow\uparrow}(ij,\tau\tau') \nn &&
\approx
\langle{T}_{\tau}[R_{i\tau}R_{j\tau'}\eta_{i\tau+}\eta_{j\tau'+}^{\dagger}]\rangle
+
\langle{T}_{\tau}[I_{i\tau}I_{j\tau'}\eta_{i\tau-}^{\dagger}\eta_{j\tau'-}]\rangle
\nn && =
\langle{T}_{\tau}[R_{i\tau}R_{j\tau'}]\rangle(\langle{T}_{\tau}[\eta_{i\tau+}\eta_{j\tau'+}^{\dagger}]\rangle
+ \langle{T}_{\tau}[\eta_{i\tau-}^{\dagger}\eta_{j\tau'-}]\rangle)
. \nn \eqa Interestingly, this structure is essentially the same
as that of the electron green's function in the slave-rotor
representation.\cite{Green_function} Since there is a gap in the
boson spectrum, the electron spectral function shows no sharp peak
at zero energy\cite{Florens} in the metallic phase, implying a
pseudogap in electron excitations. Although there exists a
pseudogap in the electron spectrum, this phase is identified with
an anomalous metal because charge carriers are not electrons but
fractionalized electrons as shown in Eq. (18).\cite{Gauge_metal}

One cautious person may suspect the stability of this metallic
phase against instanton excitations of compact U(1) gauge fields
beyond the mean field approximation.\cite{Kim_DQCP,Kim_SL_RG} The
effective U(1) gauge Lagrangian for the non-Fermi liquid metal can
be considered as \bqa && L_{eff} \approx
\sum_{i}\eta_{i}^{\dagger}(\partial_{\tau}\mathbf{I} -
ia_{1i\tau}\tau_{1})\eta_{i} \nn && -
tF\sum_{ij}\eta_{i}^{\dagger}e^{i{a}_{1ij}{\tau}_{1}}\tau_{3}\eta_{j}
- \frac{1}{g^2}\sum_{\Box}\cos(\partial\times{a}_{1}) , \eqa where
$g$ is an internal gauge charge of the $\eta_{i}$ fermions.
Condensation of the $\phi_{3i}$ bosons would result in massive
gauge fluctuations $a_{3ij}$ (Anderson-Higgs mechanism), thus
ignored in the low energy limit. It was argued that if $\tau_{1}$
in the gauge coupling is replaced with $\tau_{3}$, the effective
U(1) gauge Lagrangian lies in a critical phase identified with a
renormalization group fixed point, and can be stable against
instanton excitations due to critical matter
fields.\cite{Kim_SL_RG} However, in the present case the stability
of the anomalous metallic phase is not clearly guaranteed yet
because screening of an internal gauge charge appears in the
particle-particle channel instead of the particle-hole channel
shown in the previous study,\cite{Kim_SL_RG} thus the structure of
renormalization group equations can be totally different from
those in the previous study.

The stability of this non-Fermi liquid metal against disorder
should be also answered for this phase to be a genuine metallic
phase. The present author investigated the role of disorder in the
two dimensional fermion system with long range gauge interactions,
where $\tau_{1}$ in the gauge coupling is replaced with $\tau_{3}$
like the above.\cite{Kim_exponent,Kim_Disorder} In this case the
long range interactions are shown to make the fermion system
stable against weak disorder even in two dimensions because the
gauge interactions let the fermions lie in a critical phase.
Remember that criticality can protect fermions from localization
due to disorder.\cite{NFL_exponent,Kim_exponent,Kim_Disorder}
However, the present case should be addressed more clearly because
of difference of the gauge couplings.

In this paper we proposed that an anomalous metallic phase can
result from local pairing fluctuations near the Mott critical
point. We extended the U(1) gauge theory of the Hubbard
model\cite{Florens,LeeLee,Kim1} by allowing not only density
fluctuations but also local pairing excitations, and derived an
SU(2) gauge theory in terms of collective density and pairing
fluctuations with gapless fermion excitations. We showed that
there exists an interesting intermediate phase between the spin
liquid Mott insulator and Fermi liquid metal, characterized by
softening of local charge fluctuations as the Fermi liquid, but
gapping of local pairing excitations. This behavior of collective
bosons can be allowed from the Berry phase contribution for charge
fluctuations due to hole doping, associated with SU(2) symmetry
breaking. The intermediate phase turns out to be a metallic state
without electron-like quasiparticles, thus identified with an
anomalous non-Fermi liquid metal. Role of gauge fluctuations
should be addressed near future.

K.-S. Kim thanks Drs. J.-H. Han, A. Tanaka, M. Kohno, Y. Nonomura,
X. Hu, Q. Li, and A. Furusaki for helpful discussions.


\begin{thebibliography}{9}
\bibitem{Florens} S. Florens and A. Georges, Phys. Rev.
B {\bf 70}, 035114 (2004).
\bibitem{LeeLee} S.-S. Lee and P. A. Lee, Phys. Rev. Lett. {\bf
95}, 036403 (2005).
\bibitem{Review1} E. Abrahams, S. V. Kravchenko, and M. P.
Sarachik, Rev. Mod. Phys. {\bf 73}, 251 (2001).
\bibitem{Review2} A. Kapitulnik, N. Mason, S. A. Kivelson, and S.
Chakravarty, Phys. Rev. B {\bf 63}, 125322 (2001).
\bibitem{NFL_exponent} S. Chakravarty, L. Yin, and E. Abrahams, Phys.
Rev. B {\bf 58}, R559 (1998).
\bibitem{Kim_exponent} Ki-Seok Kim, Phys. Rev. B {\bf 72}, 014406
(2005); Phys. Rev. B {\bf 70}, 140405(R) (2004).
\bibitem{Dissipation_Metal} P. Phillips and D. Dalidovich, Phys.
Rev. B {\bf 65}, 081101 (2002); R. L. Jack and D. K. K. Lee, Phys.
Rev. B {\bf 66}, 104526 (2002); D. Das and S. Doniach, Phys. Rev.
B {\bf 60}, 1261 (1999).
\bibitem{Gradient_expansion} D. Belitz and T. R. Kirkpatrick,
Phys. Rev. B {\bf 56}, 6513 (1997); D. Belitz, T. R. Kirkpatrick,
and T. Vojta, Phys. Rev. B {\bf 65}, 165112 (2002); A. Chubukov,
C. Pepin, and J. Rech, Phys. Rev. Lett. {\bf 92}, 147003 (2004);
Ar. Abanov and A. Chubukov, Phys. Rev. Lett. {\bf 93}, 255702
(2004).
\bibitem{Slave_boson} The present decomposition scheme looks
somewhat similar to the slave-boson representation of the $t-J$
model. However, our decomposition scheme is nothing to do with the
slave-boson representation because the slave-boson representation
originates from solving the single occupancy constraint in the
$t-J$ model while the present decomposition does not. It should be
noted that this decomposition is not arbitrary since the
collective mode $U_{\sigma\sigma'}$ is determined by its HS field,
as will be discussed in the text.
\bibitem{Kim1} Ki-Seok Kim, cond-mat/0510564.
\bibitem{Canonical_method} One can derive the pair-rotor theory
Eq. (12) in the canonical quantization method by decomposing the
electron Hilbert space into the composite Hilbert space of
$|\psi\rangle = |\eta\rangle\bigotimes|\Delta^{R}\rangle$ based on
the composite field representation $\psi_{i} =
e^{-i\phi_{1i}\tau_{1}}\eta_{i}$, where $\Delta^{R}$ represents
the density of an electron pair. This enlarged composite Hilbert
space is reduced to the original electron one by the constraint
$\Delta_{i}^{R} = \psi_{i}^{\dagger}\tau_{1}\psi_{i}$. This
canonical quantization can be realized in the following path
integral expression \bqa && Z = \int{[D\eta_{i}, \phi_{1i},
\Phi_{i}^{R}, \Delta_{i}^{R}]} \exp\Bigl[ -\int{d\tau}\Bigl\{
\sum_{i}\eta_{i}^{\dagger}\partial_{\tau}\eta_{i} \nn && -
t\sum_{ij}\eta_{i}^{\dagger}e^{i\phi_{1i}\tau_{1}}\tau_{3}e^{-i\phi_{1j}\tau_{1}}\eta_{j}
\nn && + \sum_{i}\Bigl(u\Delta_{i}^{R2} -
i\Delta_{i}^{R}\partial_{\tau}\phi_{1i} + i
\Phi_{i}^{R}(\Delta_{i}^{R} - \psi_{i}^{\dagger}\tau_{1}\psi_{i})
\Bigr)\Bigr\} \Bigr] , \nonumber \eqa resulting in the pair-rotor
theory Eq. (12) after integrating out the $\Delta_{i}^{R}$ field,
and performing the HS transformation for the hopping term. The
term $- i\Delta_{i}^{R}\partial_{\tau}\phi_{1i}$ leads us to
identify the $e^{-i\phi_{1i}\tau_{1}}$ operator with an
annihilation operator of an electron pair.
\bibitem{MF_analysis} Condensation of the $\phi_{1i}$ bosons
($R_{i}$ and $I_{i}$) is determined by the following
self-consistent equations \bqa && 1 =
\int_{-D}^{D}d\epsilon{D}(\epsilon)\frac{1}{\beta}\sum_{\omega_{n}}\frac{1}{\omega_{n}^{2}/u
+E\epsilon + \lambda} , \nn && DF = -
\int_{-D}^{D}d\epsilon{D}(\epsilon)\frac{1}{\beta}\sum_{\omega_{n}}\frac{\epsilon}{\omega_{n}^{2}/u
+E\epsilon + \lambda} , \nn && DE =
-2\int_{-D}^{D}d\epsilon{D}(\epsilon)\epsilon{n}_{F}(F\epsilon) ,
\nonumber \eqa where $D$ is the half bandwidth, $D(\epsilon)$ the
band density of states, and $n_{F}(\epsilon)$ the Fermi-Dirac
distribution function. Here $-i\lambda_{i}$ is replaced with
$\lambda$. Solving the above equations with the density of states
$D(\epsilon) = 1/(2D)$ gives at zero temperature \bqa && \lambda =
\frac{u}{2} + \frac{D^2}{8u} , \nn && F =
\frac{4u^2+D^2}{16D^2u}(2u+D-\frac{5}{3}|2u-D|) , \nn && E =
\frac{1}{2} . \nonumber \eqa Condensation of the $\phi_{1i}$
bosons occurs when $\lambda - ED = 0$. These equations are exactly
the same as those for the $\phi_{3i}$ bosons as a result of the
SU(2) symmetry.
\bibitem{Magnetic_Field} This is exactly the same phenomenon as
the superconducting transition temperature decreases under
external magnetic fields since the condensation amplitude becomes
weak due to external magnetic fields. The Ginzburg-Landau
functional under external magnetic fields is given by \bqa &&
S_{GL} = \int{d^2x}\Bigl[|(\nabla - i\mathbf{A})\Phi|^{2} +
V(|\Phi|) + \frac{1}{8\pi}|\nabla\times\mathbf{A}|^{2} \Bigr] ,
\nonumber \eqa where $\Phi$ is a boson field, $V(|\Phi|)$ its
effective potential, and $\mathbf{A}$ an electromagnetic field. In
the present problem $\Phi$ corresponds to a $\phi_{3}$ vortex
field, and the external magnetic field the chemical potential,
i.e., $\nabla\times{\mathbf{A}} = \mu\mathbf{z}$. This functional
is well analyzed in the textbook, M. Tinkham, \textit{Introduction
to Superconductivity} (2nd edition), \textit{Chs. 4 and 5}
(McGraw-Hill, Inc., 1996). One can find that the presence of
magnetic fields weakens the condensation amplitude as $|\Phi|^{2}
\approx 1 - const.\cdot{H}^{2}$, where $H$ is an external magnetic
field, here $H = \mu = u\delta$.
\bibitem{Gauge_metal} It is important to realize that the dc
conductivity $\sigma_{el}$ is given by only the fermion
contribution $\sigma_{el} = \sigma_{\eta}$ instead of the
Ioffe-Larkin form [L. B. Ioffe and A. I. Larkin, Phys. Rev. B {\bf
39}, 8988 (1989)] $\sigma_{el} =
\sigma_{\phi_{1}}\sigma_{\eta}/(\sigma_{\phi_{1}} +
\sigma_{\eta})$, where $\sigma_{\phi_{1}}$ ($\sigma_{\eta}$) is
the $\phi_{1i}$ ($\eta_{i}$) conductivity, even if U(1) gauge
fluctuations $a_{1\mu}$ are allowed at the gaussian order. Note
that the $a_{3\mu}$ fluctuations are massive owing to the
Anderson-Higgs mechanism, as discussed in the text. $\sigma_{el} =
\sigma_{\eta}$ originates from the fact that there are $\tau_{1}$
and $\tau_{3}$ matrices in the gauge couplings with $a_{1\mu}$ and
$A_{\mu}$ respectively, and the correlation function between a
pair current with $\tau_{1}$ and a charge current with $\tau_{3}$
vanishes due to orthogonality between the $\tau_{1}$ and
$\tau_{3}$ matrices. As a result, the conductivity is given by the
fermion contribution only, identifying this phase with a metal
owing to the gapless $\eta_{i}$ excitations.
\bibitem{Green_function} In the slave-rotor
representation\cite{Florens} the electron propagator is given by
\bqa && G_{el\uparrow\uparrow}(ij,\tau\tau') =
\langle{T}_{\tau}[c_{\uparrow{i}\tau}c_{\uparrow{j}\tau'}^{\dagger}]\rangle
\nn && =
\langle{T}_{\tau}[e^{-i\phi_{3i\tau}\tau_{3}}\eta_{i\tau}\eta_{j\tau'}^{\dagger}e^{i\phi_{3j\tau'}\tau_{3}}]_{11}\rangle
\nn && =
\langle{T}_{\tau}[e^{-i(\phi_{3i\tau}-\phi_{3j\tau'})}\eta_{i\tau+}\eta_{j\tau'+}^{\dagger}]\rangle
\nn && +
\langle{T}_{\tau}[e^{i(\phi_{3i\tau}-\phi_{3j\tau'})}\eta_{i\tau-}^{\dagger}\eta_{j\tau'-}]\rangle
\nn && \approx
\langle{T}_{\tau}[e^{i(\phi_{3i\tau}-\phi_{3j\tau'})}]\rangle(\langle{T}_{\tau}[\eta_{i\tau+}\eta_{j\tau'+}^{\dagger}]\rangle
+ \langle{T}_{\tau}[\eta_{i\tau-}^{\dagger}\eta_{j\tau'-}]\rangle)
. \nonumber \eqa
\bibitem{Kim_DQCP} Ki-Seok Kim, Phys. Rev. B {\bf 72}, 035109
(2005); Phys. Rev. B {\bf 72}, 214401 (2005).
\bibitem{Kim_SL_RG} Ki-Seok Kim, Phys. Rev. B {\bf 72}, 245106 (2005).
\bibitem{Kim_Disorder} Ki-Seok Kim, Phys. Rev. B {\bf 73}, 235115
(2006).
\end{thebibliography}
\end{document}